\documentclass[
prx,
 amsmath,amssymb,
 superscriptaddress,
 reprint,%
]{revtex4-2}

\usepackage{graphicx}
\usepackage{dcolumn}
\usepackage{bm}
\usepackage{hyperref}

\usepackage[utf8]{inputenc}
\usepackage[T1]{fontenc}
\usepackage{etoolbox}
\usepackage{xcolor}

\usepackage{float}
\usepackage{amsfonts}
\usepackage{braket}

\def\sn{^{(n)}}

\newcommand{\Google}{\affiliation{Google Quantum AI, Goleta, California 93117, USA}}

\begin{document}
\begin{title}


\title{Josephson parametric amplifier with Chebyshev gain profile and high saturation}

\author{Ryan Kaufman}\Google
\affiliation{Department of Physics and Astronomy, University of Pittsburgh, Pittsburgh, Pennsylvania 15260, USA}
\author{Theodore White}\Google
\author{Mark I. Dykman}
\affiliation{Michigan State University, East Lansing, MI 48823, USA}
\author{Andrea Iorio}\Google
\affiliation{NEST, Istituto Nanoscienze-CNR and Scuola Normale Superiore, I-56127 Pisa,
Italy}
\author{George Sterling}\Google
\author{Sabrina Hong}\Google
\author{Alex Opremcak}\Google
\author{Andreas Bengtsson}\Google
\author{Lara Faoro}\Google
\author{Joseph C. Bardin}\Google
\affiliation{Department of Electrical and Computer Engineering, University of Massachusetts, Amherst, Massachusetts 01003, USA}
\author{Tim Burger}\Google
\author{Robert Gasca}\Google
\author{Ofer Naaman}\Google

\date{\today}

\begin{abstract}
We demonstrate a Josephson parametric amplifier design with a band-pass impedance matching network based on a third-order Chebyshev prototype. We measured eight amplifiers  operating at 4.6~GHz that exhibit gains of 20~dB with less than 1~dB gain ripple and up to 500~MHz bandwidth. The amplifiers further achieve high output saturation powers around $-73$~dBm based on the use of rf-SQUID arrays as their nonlinear element. We characterize the system readout efficiency and its signal-to-noise ratio near saturation using a Sycamore processor, finding the data consistent with near quantum limited noise performance of the amplifiers. In addition, we measure the amplifiers' intermodulation distortion in two-tone experiments as a function of input power and inter-tone detuning, and observe excess distortion at small detuning with a pronounced dip as a function of signal power, which we interpret in terms of power-dependent dielectric losses. 
\end{abstract}

\maketitle
\end{title}

\section{Introduction}\label{sec:intro}
Josephson parametric amplifiers (JPAs) \cite{aumentado2020superconducting} are critical components in superconducting quantum computing architectures that rely on dispersive readout. They provide a first gain stage with near quantum-limited noise and, when followed by commercial low-noise cryogenic amplifiers and room-temperature receivers, enable fast and accurate detection of low power readout signals \cite{jeffrey:readout}. Today's intermediate scale superconducting quantum processors \cite{google:quantsup, krinner2022realizing, acharya2022suppressing} employ frequency domain multiplexing to readout multiple qubits with each measurement chain \cite{heinsoo2018rapid}. JPAs are therefore required to feature both high instantaneous bandwidth (at least 500~MHz), to accommodate sufficient spectral separation between readout tones, and high dynamic range (output 1~dB compression power exceeding $-90$~dBm), to avoid loss of readout fidelity due to gain compression and intermodulation distortion. These requirements have led to the development of superconducting traveling wave parametric amplifiers \cite{macklin:JTWPA, esposito2021perspective, eom:TiNparamp}, and a quest to improve the dynamic range and instantaneous bandwidth of resonator-based JPAs. 

While high dynamic range JPAs based on rf-SQUID \cite{white2022readout} and SNAIL \cite{frattini2018optimizing, sivak2019kerr} arrays have been demonstrated, achieving a reliable and predictable broadband gain has proved more difficult. Impedance matched JPAs have also been demonstrated \cite{IMPA, ranzani2022wideband}, but these derive their broadband performance and exact gain profile primarily from hard to control details in the microwave environment that are extrinsic to the amplifiers themselves. A controllable, engineered broadband response can nevertheless be achieved by using band-pass impedance matching circuits tailored for a specific gain profile \cite{vijay:broadband,ezenkova2022broadband}, harnessing network synthesis techniques common in microwave engineering \cite{naaman2022synthesis}.

Here we report on Josephson parametric amplifiers with three-pole Chebyshev matching networks, designed according to Ref.~\cite{naaman2022synthesis} to produce a broadband gain profile with controlled ripple. Unlike Refs.~\cite{IMPA, ranzani2022wideband,white2022readout}, broadband performance here is engineered from the outset, and is not an accidental consequence (however advantageous) of external factors. We use rf-SQUID arrays (`snakes') as in Ref.~\cite{white2022readout} to ensure high dynamic range, and implement the matching network using passive, on-chip, lumped element components. These Lumped Element Snake Amplifiers (LESAs) exhibit 20~dB gain with less than 1~dB ripple, up to 500~MHz bandwidth, and typical output saturation power of $-73$~dBm. We characterize eight LESA amplifiers using a 54-qubit Sycamore processor and measure their readout efficiency, as well as gain and noise compression. 

We additionally measured the LESA intermodulation distortion in two-tone experiments, where we observe a surprising effect\textemdash excess intermodulation distortion with nontrivial dependence on both tone power and inter-tone spacing. We explain this effect by considering nonlinear dielectric losses in the amplifier due to a saturable bath of two-level system defects.

\section{Design}\label{sec:design}
The devices were designed for a center frequency of $\omega_0/2\pi=4.9$~GHz and a fractional bandwidth of 0.135 ($\Delta\omega/2\pi = 660$~MHz), using a 20 dB gain, 0.5 dB ripple third-order Chebyshev prototype \cite{naaman2022synthesis} 
\begin{equation}\label{eq:coeffs}
\left\{g_0,\dots,g_4\right\}=\left\{1.0,\; 0.5899,\;0.6681,\;0.3753,\;0.9045\right\}.
\end{equation}
The coefficients $g_i$ in Eq.~(\ref{eq:coeffs}) are normalized conductances of the low-pass ladder network prototype, and relate to the polynomials defining the input impedance of the network as a function of frequency \cite{pozar2009microwave}. Coefficient $g_0$ corresponds the parametrically pumped inductance, and coefficient $g_4$ corresponds the impedance of the load. 

\begin{figure}[ht]
\includegraphics[width=\columnwidth]{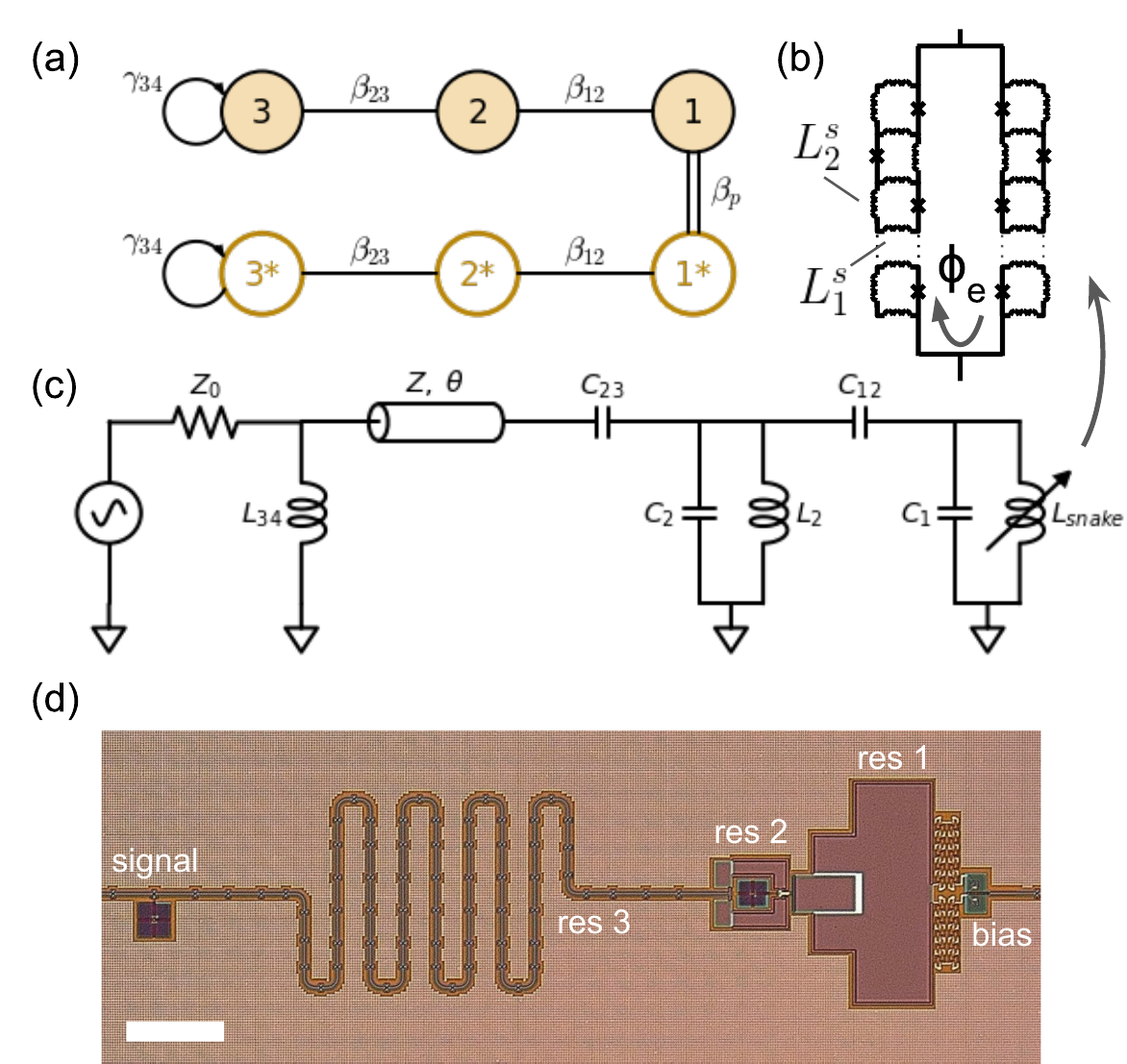}
\caption{\label{fig:schem} (a) Coupled mode graph of the device, filled- and open-face shading indicate co-rotating (signal) and conjugate (idler) modes, respectively. Modes 1 and $1^*$ are coupled parametrically, and the rest of the couplings are passive. (b) Schematic of the `snake' rf-SQUID array nonlinear element, modeled as variable inductance $L_\mathrm{snake}$ tunable via an applied flux $\phi_e$ from an on-chip superconducting transformer. (c) Circuit schematic of the LESA amplifier, and (d) an optical micrograph of the device. The signal port is on the left, and the snake arrays, bias transformer, and pump line are on the right. Scale bar is 120~$\mu$m. }
\end{figure}

Figure~\ref{fig:schem}(a) shows the coupled mode graph \cite{ranzani2015graph, naaman2022synthesis} of the device, having three co-rotating modes with frequency $\omega_0$ (filled face in the figure) and three corresponding conjugate modes (open face). The corresponding matrix that encapsulates the equations of motion for the mode amplitudes is given in Appendix~\ref{apx:matrix}. This is a degenerate parametric amplifier, so both co-rotating and conjugate modes are hosted within the same physical three-resonator circuit.  Resonator 3 is coupled to the $Z_0=50\,\Omega$ environment with a dissipation rate of 
\begin{equation}
\gamma_{34}/2\pi = \frac{\Delta\omega}{2\pi g_3g_4}=1.95\,\mathrm{GHz};
\end{equation}
this is also the characteristic decay rate \cite{naaman2022synthesis} $\gamma_0\equiv\gamma_{34}$ used below. Resonator 2 is coupled passively to both resonator 1 and 3, with reduced coupling rates
\begin{align}
    \beta_{23} & = \frac{\Delta\omega}{2\gamma_0\sqrt{g_2g_3}} = 0.339 \label{eq:beta23}\\
    \beta_{12} & = \frac{\Delta\omega}{2\gamma_0\sqrt{g_1g_2}} = 0.270\label{eq:beta12}.
\end{align}
The strength of the parametric coupling between mode 1 and its conjugate is \cite{naaman2022synthesis}
\begin{equation}
    \beta_p = \frac{1}{2}\frac{g_3g_4}{g_0g_1} = 0.288.
\end{equation}

Because of variations in the fabrication process and other uncertainties, we can only know component values to within $\pm10\%$. The measured devices operate at a lower center frequency of 4.6~GHz and with a smaller bandwidth than designed. In the following, we will report the nominal component values for the design, keeping in mind the uncertainty in their final `as fabricated' values.

The circuit schematic of the LESA is shown in Fig.~\ref{fig:schem}(c). Resonator 1 is formed by a capacitor $C_1 = 6.6$~pF shunting the nonlinear snake inductance $L_\mathrm{snake}$. The snake, shown schematically in Fig.~\ref{fig:schem}(b), is composed of two parallel arrays of $N$ rf-SQUIDs, where each rf-SQUID contains a Josephson junction with critical current $I_c$, and a linear inductance made out of two segments with inductance $L^s_1$ and one segment with inductance $L^s_2$, such that the $L^s_1$ segments are shared between neighboring SQUIDs \cite{white2022readout}. The snake used here is identical to that in Ref.~\cite{white2022readout}, with a total of $2N=40$ rf-SQUIDs, junction $I_c=16\,\mu A$, and inductances $L^s_1=2.6$~pH and $L^s_2=8.0$~pH. It is flux biased via an on-chip superconducting transformer to set resonator 1's frequency to $\omega_0$, and parametrically flux pumped at $\omega_p = 2\omega_0$. Resonator 2 is implemented as a lumped-element parallel $LC$ resonator with $C_2 = 0.65$~pF and $L_2=0.65$~nH. The coupling corresponding to $\beta_{12}$ in Eq.~(\ref{eq:beta12}) is realized by capacitor $C_{12}=0.74$~pF, and that corresponding to $\beta_{23}$ is the coupling capacitor $C_{23}=0.27$~pF. Resonator 3 is implemented as a transmission line resonator with a characteristic impedance $Z=50\,\Omega$ and an electrical length of $\theta=32.6^\circ$ at $\omega_0$, and is coupled inductively to the $50\,\Omega$ signal port with $L_{34}=1.32$~nH. All values above are calculable (see Appendix~\ref{apx:componenets}) given the prototype in Eq.~(\ref{eq:coeffs}), the center frequency, and the bandwidth of the amplifier, using standard filter design techniques \cite{naaman2022synthesis}. The electrical length of resonator 3 was further trimmed manually by $-6^\circ$ compared to its calculated value, based on results from harmonic balance circuit simulations. This is presumably needed in order to compensate for the frequency dependence of all coupling elements, which are only evaluated at $\omega_0$.

Figure \ref{fig:schem}(d) shows an optical micrograph of the LESA, with the snake inductor and bias line on the right, and the signal port on the left. The devices were built in a three layer aluminum process with SiO$_x$ interlayer dielectrics and Al/AlO$_x$/Al trilayer Josephson junctions.

\section{Gain, Saturation, and Readout Efficiency}\label{sec:gain}
Eight LESA devices were packaged in magnetically shielded enclosures and mounted on the mixing chamber of a dilution refrigerator hosting a 54-qubit Sycamore processor \cite{google:quantsup}. Each amplifier connects to one of the processor's readout line (labeled A-G, I) via four circulators. The LESA associated with lines A-E differ (by $\approx10\%$) from those on lines F, G, and I, in the width of the center conductor of resonator 3 but with no discernable effect on their performance. Readout line H was outfitted with a standard dc-SQUID based IMPA \cite{IMPA}. 

\begin{figure}[th]
\includegraphics[width=3.2in]{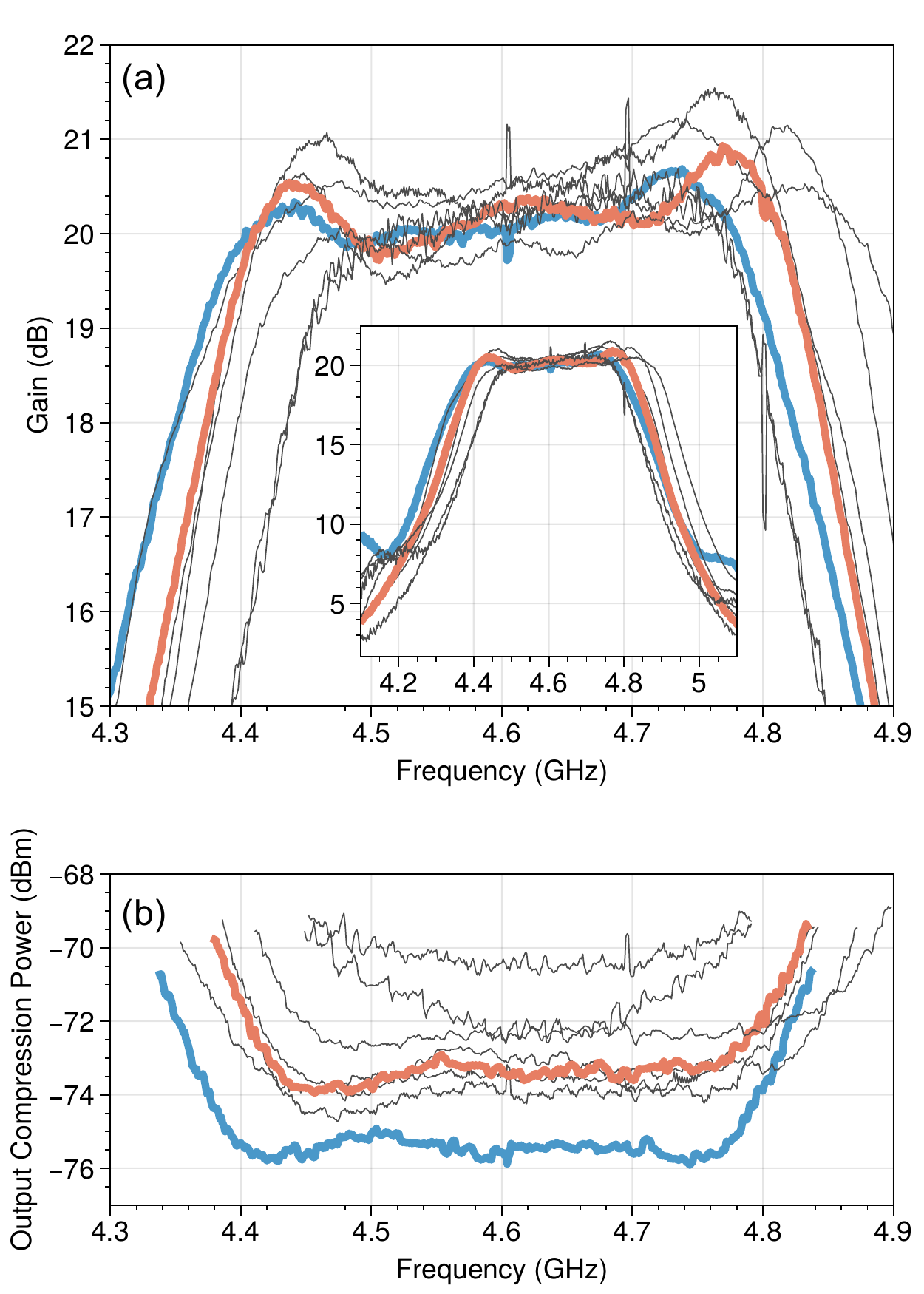}
\caption{\label{fig:gain_sat} (a) Gain in dB vs frequency of all LESA devices after manual tuneup. Inset: same data plotted over a wider range of frequency and gain. (b) Output 1-dB compression power. Power was calibrated at a reference plane at the input to the amplifier, with uncertainty of $\pm 1$~dB. In both panels, readout line A is highlighted in blue, and readout line F in orange.}
\end{figure}

The amplifiers' flux biases, pump powers, and pump frequencies were tuned manually to optimize their bandwidth while maintaining 20 dB gain with at most 1 dB ripple. Figure~\ref{fig:gain_sat}(a) shows the resulting gain vs signal frequency of all LESA, highlighting the ones on readout line A (blue) and F (orange). The inset shows the same data plotted over a wider range of frequency and gain.

Figure~\ref{fig:gain_sat}(b) shows the output saturation power (1-dB gain compression) measured vs signal frequency for all LESA using the same operating point as in (a). Power was calibrated by measuring the ac-Stark shift for each of the qubits on each of the readout lines at the readout resonators' dressed frequencies \cite{white2022readout}. The average power over all qubits in a readout line was then used to calibrate the room-temperature generator power to that at a reference plane on the processor chip. That calibration was then transferred to a reference plane at the input of the LESA by including independently measured losses between the processor and the LESA. Overall, accounting for frequency-dependent variation in the ac-Stark calibration and uncertainties in the loss estimates, the uncertainty in the power calibration is $\pm 1$~dB. The Figure shows that the typical output saturation power is around $-73$~dBm, corresponding to an input saturation power of $\mathrm{IP}_{1\mathrm{dB}}=-93$~dBm at 20~dB gain. These saturation powers agree with those reported in Ref.~\cite{white2022readout}, and represent roughly 100-fold increase (20~dB) over typical values for single dc-SQUID based JPAs. The change in the phase of the amplified signal at the 1-dB compression point is less than $5^\circ$ compared to its low power value.

\begin{figure}
\includegraphics[width=\columnwidth]{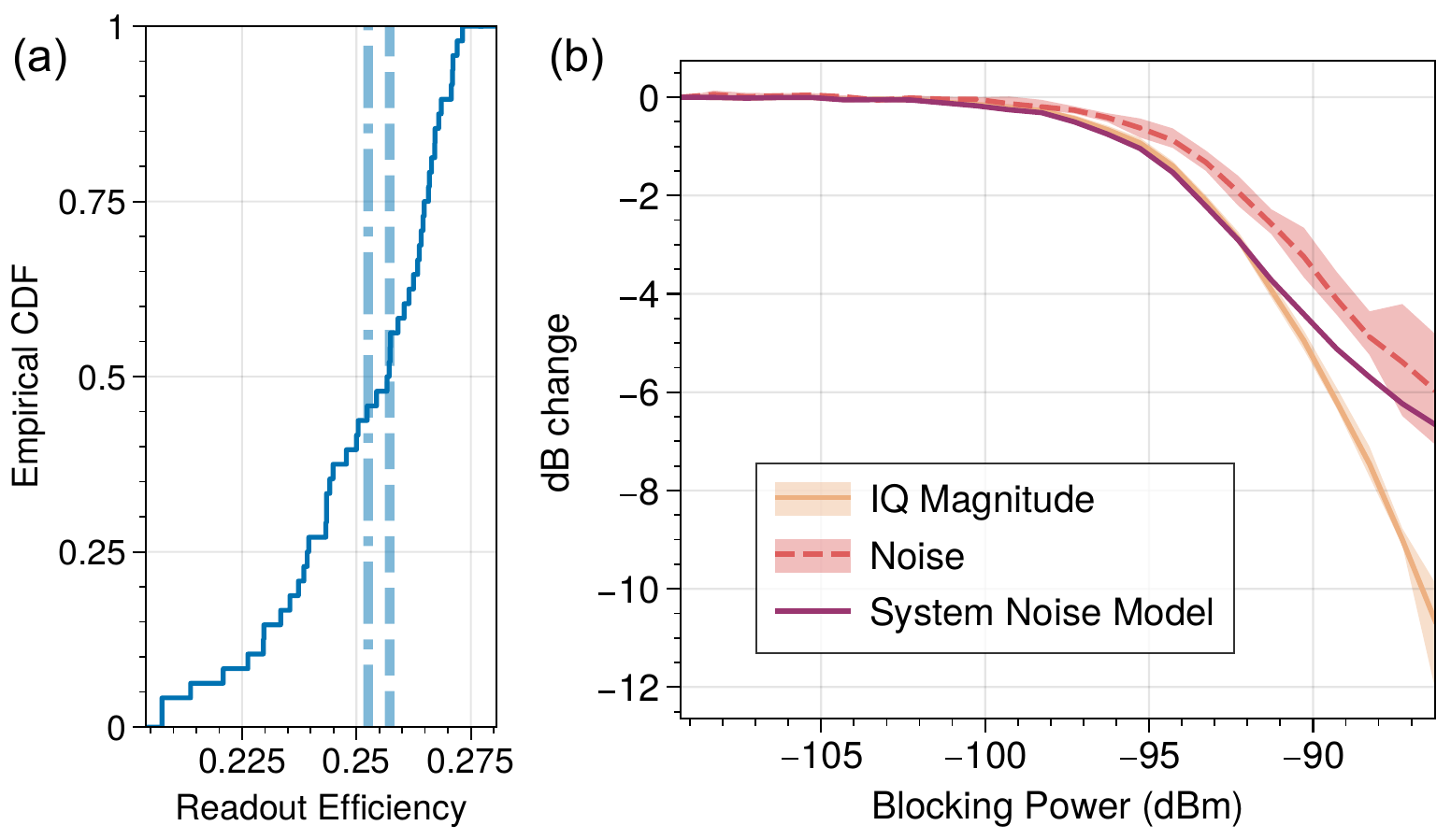}
\caption{\label{fig:efficiency_noise} (a) Readout efficiency empirical cumulative distribution function (CDF), with a median of 0.257  (dashed), and a mean of  0.253 (dash-dot).  (b) Normalized readout cloud magnitude (yellow) and noise (pink) in dB, measured on readout line A in the presence of a blocking tone, vs blocking tone power. Solid lines represent the mean over qubits, and shading represents the range of the data. Purple - system noise model accounting for reduction of the SNR due to gain compression alone.}
\end{figure}

Figure~\ref{fig:efficiency_noise}(a) shows the empirical cumulative distribution function of the readout efficiency, measured similarly to Ref.~\cite{white2022readout}, on all eight readout lines outfitted with LESAs. The median value of 0.257 (the maximum possible efficiency is 0.5) is consistent with near quantum limited noise performance of the LESAs if the average microwave loss between the processor and the amplifiers is $-1.9$~dB. Independent estimates of these losses from cryogenically calibrated measurements \cite{wang2021cryogenic, ranzani2013two} of individual components and integrated assemblies are between $-1.75$~dB and $-1.95$~dB at 4.6 GHz. The lower losses here, compared to Ref.~\cite{white2022readout}, are a result of deliberate improvements in our readout assembly. 

To investigate how the readout signal-to-noise ratio (SNR) degrades when the LESAs are driven to saturation, we performed `readout clouds' measurements with qubits on readout line A in the presence of a blocking tone\textemdash an additional tone at 4.4~GHz, near the edge of the amplifier band, whose purpose is to saturate the amplifier. Qubits on readout line A (except for one, whose readout frequency coincided with the idler of the blocking tone) were prepared in either the $|0\rangle$ or $|1\rangle$ states, and the demodulated readout signal (in-phase and quadrature, IQ) for each qubit was recorded. When repeated over many shots, this measurement produces two point-clouds in the IQ plane (a symbol constellation in digital communications nomenclature)  corresponding to the two prepared states of the qubit. The separation between the clouds, which is proportional to the magnitude of the IQ vector from the origin to the centers of the clouds (`IQ magnitude' below), is the signal in this measurement; the clouds' standard deviation is a measure of the noise.

The change in the received signal, the readout clouds' IQ magnitude, is shown in yellow in Figure~\ref{fig:efficiency_noise}(b), and the change in noise is shown in pink, as a function of blocking tone power. The solid curves show the average, and the shading represents the range of the data, over the five measured qubits. We see that the signal magnitude degrades as expected when the blocking tone power reaches the LESA input saturation point. As the gain of the LESA decreases near saturation, the measured output noise is expected to decrease as well, assuming a constant input noise power, and then level off as noise contribution from the cryogenic HEMT amplifier becomes more dominant. The expected change in the output system noise power due to LESA gain compression alone is shown in purple in the figure (`system noise model'), assuming quantum limited LESA noise and a HEMT noise temperature of 2.5~K. The measured noise (pink, dashed) clearly falls above the system noise model prediction, or in other words, the signal compresses before the noise does. We note, however, that we do not observe a noise peaking phenomenon such as reported in Ref.~\cite{remm2022intermodulation} for a 4-wave mixing Josephson traveling wave amplifier.

\section{Intermodulation Distortion}\label{sec:IMD}
\begin{figure}[ht]
\includegraphics[width=\columnwidth]{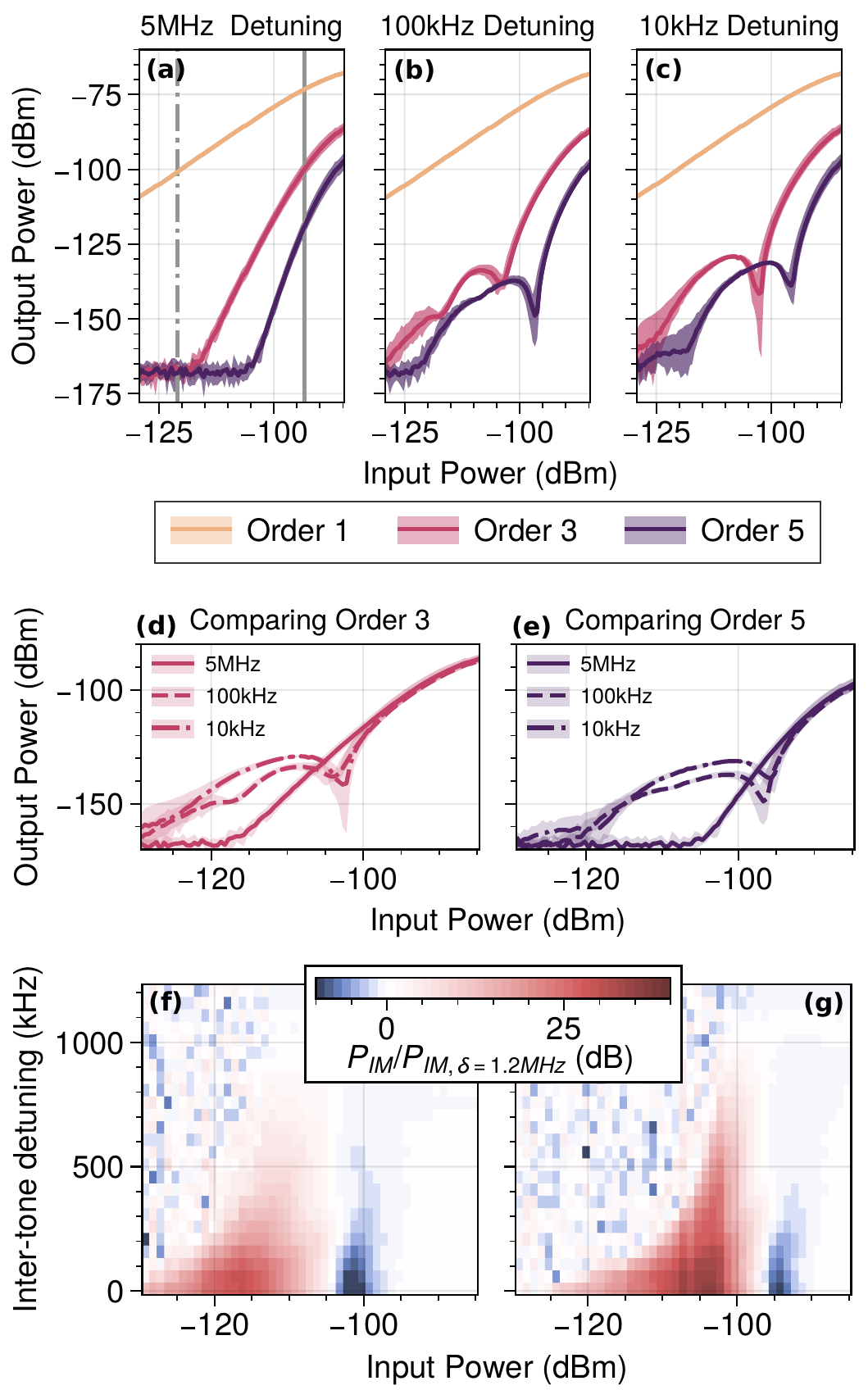}
\caption{\label{fig:imd} (a)-(c) fundamental $f_1$ (yellow, order 1), order 3 IM at $2f_1-f_2$ (pink), and order 5 IM at $3f_1-2f_2$ (purple), in a two-tone experiment with $f_1$ and $f_2$ separated by (a) $\Delta f=5$~MHz, (b) $\Delta f=100$~kHz, and (c) $\Delta f=10$~kHz. The center frequency $f_c=(f_1+f_2)/2$ is swept across the amplifier band, solid lines represent the mean, and shaded areas represent the range over the center frequency sweep. Solid gray line in (a) indicates input 1-dB saturation power, dash-dot line indicates a typical total readout pulse power. (d) Comparison of 3rd order IM for different inter-tone detunings, (e) Comparison of 5th order IM for different inter-tone detunings. (f)-(g) excess IM product power, in dB, relative to large detuning IM power (referenced at 1.2 MHz detuning) as a function of input power and inter-tone detuning. (f) represents order 3 and (g) represents order 5 IM.}
\end{figure}

Next, we turn to characterizing the LESA intermodulation distortion \cite{pozar2009microwave}. The experiments, shown in Figure~\ref{fig:imd}, were performed on readout line F, by combining two tones from two independent signal generators (using a Wilkinson power combiner) and feeding them into the readout line input. The tones had nominally the same power, shown on the x-axes in the figure. The tone frequencies, $f_1$ and $f_2$, were separated by the inter-tone detuning $\Delta f$ and centered around $f_c$. The output signal was measured using a spectrum analyzer with a resolution bandwidth of 10 Hz.

Figure~\ref{fig:imd}(a) shows the single-sideband output power at the fundamental frequency $f_1$ (yellow), the third-order intermodulation (IM) product at $2f_1-f_2$ (pink), and the fifth-order IM product at $3f_1-2f_2$ (purple), for an inter-tone separation (detuning) of $\Delta f=5$~MHz. The center frequency $f_c$ was swept across the lower half of the amplifier band, from 4.4~GHz to 4.55~GHz; the solid traces in the figure represent the average, and the shading represents the range of the data over the $f_c$ sweep. The measured 1-dB input compression power (IP$_{1\mathrm{dB}}$) of this amplifier is indicated by the vertical solid gray line, and the total readout power under normal operating conditions is indicated by the vertical dash-dot line. We see that at this detuning, intermodulation distortion (IMD) follows the expected behavior \cite{pozar2009microwave, frattini2018optimizing, remm2022intermodulation} (as is also the case with higher order IM products, not shown) with the appropriate IM product slopes vs input power. At typical readout powers the IM products fall below $-60$~dBc. We note that in multiplexed readout, the typical frequency separation between simultaneously applied readout tones is several tens of MHz \cite{google:quantsup,krinner2022realizing}, so the data in Fig.~\ref{fig:imd}(a) are representative of what could be expected in this context.

Figure~\ref{fig:imd}(b) and \ref{fig:imd}(c) show the results of the experiment with smaller inter-tone detunings, $\Delta f=100$~kHz and $\Delta f=10$~kHz respectively. Here, we see unexpected excess IMD, which is nonmonotonic with input power, and with features that disperse as a function of detuning $\Delta f$, but depend only weakly on $f_c$. In particular, the IM power at all orders exhibit a pronounced dip as a function of input power. Figures~\ref{fig:imd}(d) and \ref{fig:imd}(e) directly compare the third-order and fifth-order IMD, respectively, for different detunings. 

Figures~\ref{fig:imd}(f) and \ref{fig:imd}(g) show, as a function of input power and inter-tone detuning, the excess 3rd- and 5th-order IM power, respectively. The data here was normalized by the respective IM power at a $
\Delta f=1.2$~MHz, which is representative of the `large detuning' response that is dominated by the amplifier Kerr nonlinearity. Data from LESA on all readout lines are in qualitative mutual agreement and reproducible with several variations of the experimental setup. We therefore look for a physical mechanism associated with the device itself.

Anomalous IMD that depends on the inter-tone detuning is well documented in semiconductor microwave and power amplifiers \cite{brinkhoff2003effect,de2002comprehensive, le2001analysis}, as well as in passive microwave structures with temperature or power dependent material properties \cite{hein2002nonlinear, ott2004nonlinear, rocas2010passive}. The voltage waveform of a 2-tone drive can be written as $V(t)=V_d\left(\cos\omega_1t+\cos\omega_2t\right)=2V_d\cos{(\Delta\omega t/2)}\cos{(\omega_ct)}$, meaning that the instantaneous power at the the center frequency $\omega_c=2\pi(f_1+f_2)/2$ is slowly modulated at the rectified beat frequency $\Delta\omega=2\pi(f_2-f_1)$. In a power-dependent medium, this modulation can mix with the signal and generate a product at $2f_1-f_2$. In our devices, which are built with SiO$_x$ interlayer dielectrics and operate at mK temperatures and low power levels, the power-dependent loss tangent of the dielectrics due to a bath of two-level systems \cite{mcdermott2009materials, sage2011study, faoro2015interacting, phillips1987two} can be responsible for the observed nonlinearity. Since the two-level system (TLS) relaxation-saturation dynamics are not instantaneous, this process depends on the inter-tone detuning as observed, favoring low beat frequencies (small detuning) and rejecting faster modulation (large detuning), on a characteristic scale of $T_2$, the TLS dephasing time. The TLS-induced nonlinearity has the opposite sign with respect to the usual softening Kerr nonlinearity of the amplifier, giving rise to the prominent dip feature seen in Fig.~\ref{fig:imd}(b)-(e) when the contributions from the two nonlinear process cancel. 

\begin{figure}
\includegraphics[width=\columnwidth]{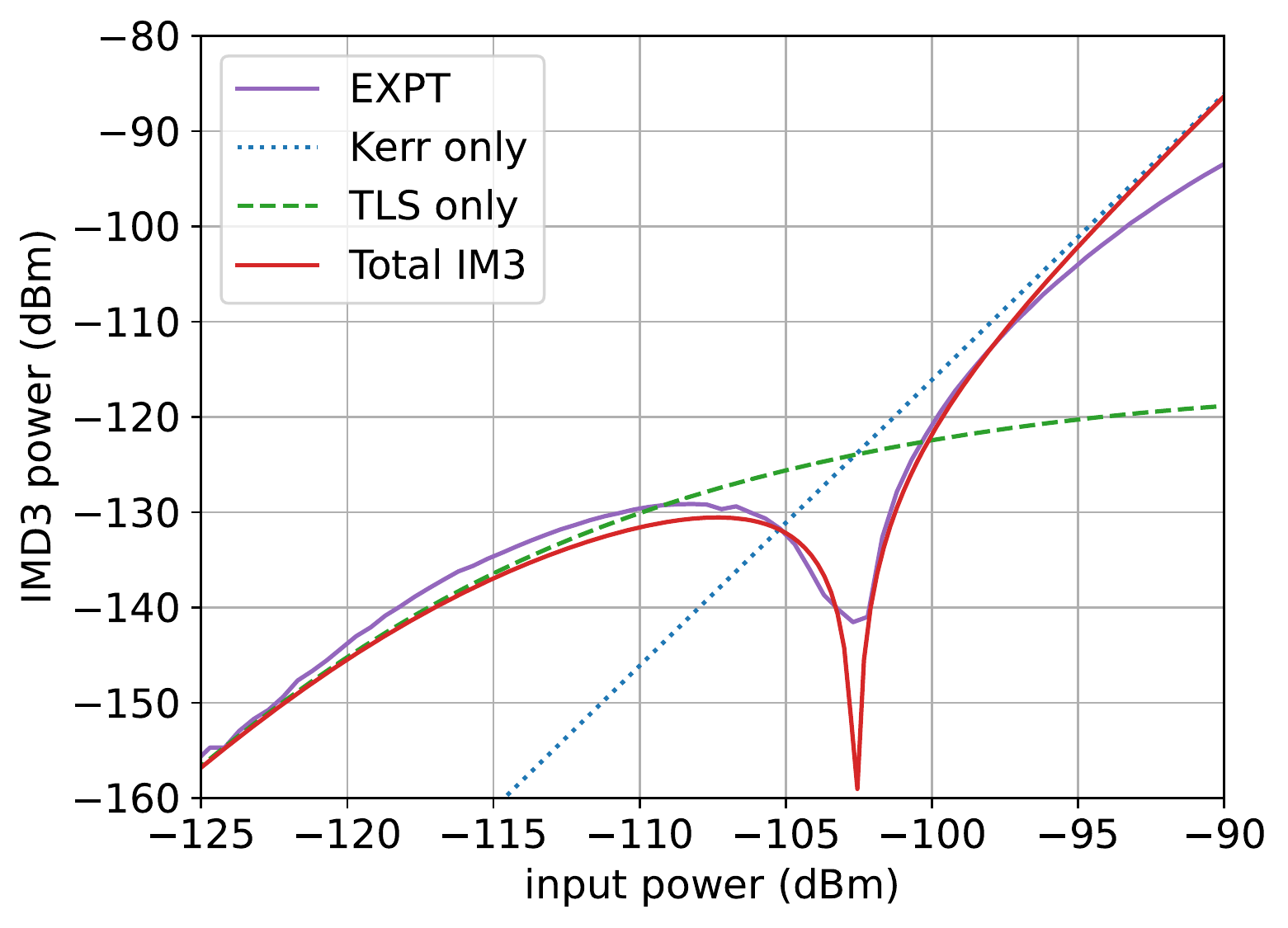}
\caption{\label{fig:TLS} Measured 3rd-order IM product at 10~kHz inter-tone detuning (purple), averaged over $f_c$ in the lower half of the amplifier band, compared with the calculated IM signal based on Eq.~(\ref{eq:thy_IMD}), with $T_1=2\,\mu$s, $T_2=2T_1$, $Q_i=250$, $G=20$~dB, $w=0.085$, $Z_1=4.4\,\Omega$, and $K_3=2.1\times10^{-3} \mu V^{-2}$ estimated from the amplifier saturation power. The Rabi frequency $\bar{\Omega}_R$ was calculated at each input power assuming a TLS dipole moment of 1 debye. The individual contributions from TLS and Kerr nonlinearities are shown in green (dashed) and blue (dotted), respectively.}
\end{figure}

We solve the Bloch equations for TLSs \cite{phillips1987two} resonant with $\omega_c$ to find their polarization under a two-tone drive, and then calculate the response of the system at the third order IM product frequency $2f_1-f_2$ (see Appendix~\ref{apx:imd_thy}). We focus on small inter-tone detuning, $\Delta\omega\ll 1/T_2$, so that the TLSs follow the beat envelope adiabatically. Considering only contributions from TLSs in the LESA primary capacitor $C_1$, we can write the IM product as
\begin{equation}\label{eq:vtls}
    V_\mathrm{TLS} = \frac{3G}{4\pi Q_i}\frac{\omega_0V_d}{\kappa T_1\bar{\Omega}_R^2(V_d)}\langle \Psi^{2\omega_1-\omega_2} \rangle,
\end{equation}
where $G$ is the amplifier power gain and $Q_i$ is the low-power internal quality factor of the LESA primary resonator. $\kappa=w\omega_0/g_1$ is the resonator external damping rate, where $w$ is the fractional bandwidth of the matching network, and $g_1$ is the filter prototype coefficient corresponding to the resonator, whose impedance is $Z_1$. $\bar{\Omega}_R(V_d)$ is the amplitude of modulation of Rabi frequency of the TLS, driven by the amplified, slowly time varying intra-cavity field, and $T_1$ is the TLS characteristic energy relaxation time. $V_d = V_\mathrm{in}\sqrt{\frac{Z_1g_1}{wZ_0}}$ is the amplitude of the drive voltage on the LESA capacitor, where $V_\mathrm{in}$ is the amplitude of input signal, and $Z_0=50\,\Omega$. The function $\Psi^{2\omega_1-\omega_2}(\xi)$ is
\begin{equation}
   \Psi^{2\omega_1-\omega_2}(\xi) = \frac{1}{4}\sqrt{\xi+1}+\frac{3}{4}\xi^{-\frac{1}{2}}\log{\left(\sqrt{\xi}+\sqrt{\xi+1}\right)}-1,
\end{equation}
whose argument $\xi$ depends on $\delta\omega$, the detuning between the TLS resonance frequency and $\omega_c$, and the Rabi frequency $\bar{\Omega}_R(V_d)$,
\begin{equation}
    \xi = \frac{2T_1T_2\bar{\Omega}_R^2(V_d)}{1+\left(T_2\delta\omega\right)^2}.
\end{equation}
The angle brackets in Eq.~(\ref{eq:vtls}) represent averaging over all TLS detunings $\delta\omega$. Finally, the output signal at the 3rd-order IM frequency is
\begin{equation}\label{eq:thy_IMD}
    V_{out} = V_\mathrm{TLS}\sqrt{\frac{wZ_0}{g_1g_4Z_1}}-\frac{3}{4}GK_3V_{in}^3,
\end{equation}
where the second term is the usual contribution from the amplifier's Kerr nonlinearity with a coefficient $K_3$.

The contribution to the IM power due to the saturable TLS bath is shown in green (dashed) in Figure~\ref{fig:TLS} with the parameters given in the caption. At low drive powers, $V_\mathrm{TLS}$ grows like $V_{in}^3$ but then levels off as the TLS bath becomes saturated over an increasing fraction of the 2-tone beat period. When combined with the Kerr contribution (dotted, blue), the total output signal, Eq.~(\ref{eq:thy_IMD}), reproduces the main features of the experimental data at small inter-tone detuning (purple). Additional IMD features that are visible in Fig.~\ref{fig:imd} are likely due to contributions from TLSs residing in the other capacitors of the LESA. When the inter-tone detuning $\Delta\omega\gg1/T_2$, the TLS polarization cannot follow the 2-tone beat envelope: the loss becomes time-independent and no longer contributes to the IMD.

We are not aware of previous observations of dynamic TLS nonlinearity in low power, low temperature experiments. It is observed here due to a combination of factors: the relatively low quality factor of the dielectrics, and the relatively high linearity of the amplifier itself. These results point to an intriguing opportunity to use intermodulation distortion in multi-tone experiments as a tool to characterize TLS dynamics in amorphous dielectrics.

\section{Conclusion}
In conclusion, we have demonstrated Josephson parametric amplifiers that have both high output saturation powers, $\approx -73$~dBm (input saturation IP$_\mathrm{1dB}=-93$~dBm), and bandwidths of up to 500~MHz, with 20~dB of gain and less than 1~dB gain ripple. The amplifiers derive their wide bandwidth from a band-pass impedance matching network based on a Chebyshev prototype, and their high dynamic range from the use of high critical current rf-SQUID arrays as their nonlinear element. We measured readout efficiencies with a median of 0.26, consistent with near quantum limited noise performance, and investigated the readout SNR degradation near saturation using a Sycamore processor. We measured the amplifiers' intermodulation distortion and observed an unexpected anomalous excess IMD at inter-tone detunings below $\approx 1$~MHz, which we can understand in terms of power-dependent TLS losses in our dielectrics. Aside from the amplifiers' favorable performance in the context of frequency-multiplexed readout, the predictability of the gain profile presents a significant practical advantage in the amplifier bring-up procedure, enabling the use of a calculable, canonical gain curve as the target for automated optimization of pump power, pump frequency, and flux bias. 

\begin{acknowledgments}
We are grateful to the Google Quantum AI team for building, operating, and maintaining software and hardware infrastructure used in this work. We thank V.~Sivak, D.~Sank, and M. Hatridge for review of the manuscript, and A.~Korotkov for technical assistance.
\end{acknowledgments}

\begin{appendix}
\renewcommand{\thefigure}{S\arabic{figure}}%
\setcounter{figure}{0}

\section{Circuit simulations}\label{apx:simulation}
Here, we review several methods that we have used to simulate the LESA circuit: S-parameter calculation based on the inverse of the ideal coupled-mode equations-of-motion matrix \cite{naaman2022synthesis}, harmonic balance circuit simulation, and linear S-parameter circuit simulation.

\begin{figure}
\includegraphics[width=\columnwidth]{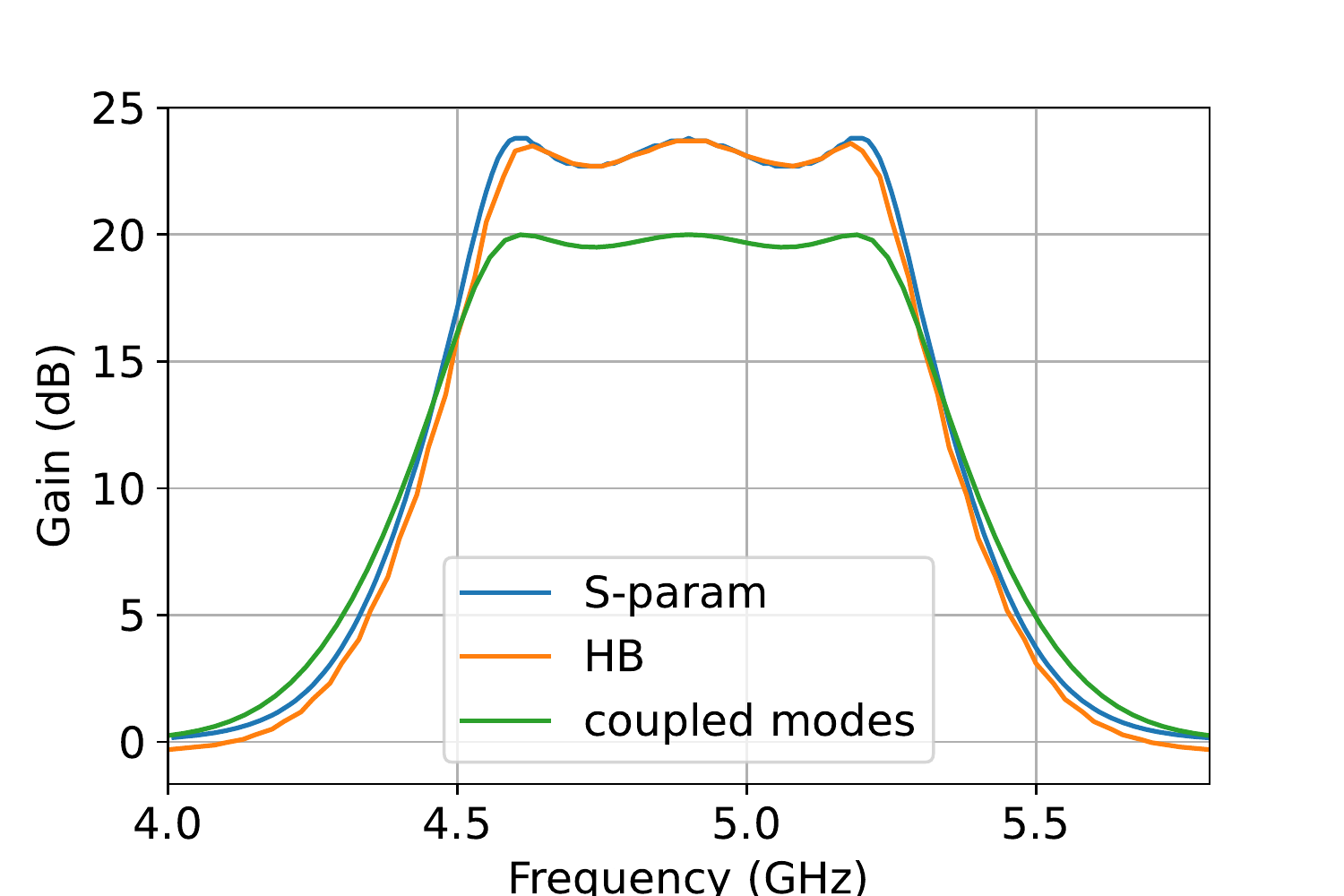}
\caption{\label{fig:sim_compare} Comparing results of simulations of the LESA using different methods. The coupled-mode simulation (green) is based on the inverse of the ideal coupled-mode matrix, while the harmonic balance (orange) and linear S-parameters (blue) are circuit simulations in Keysight ADS.}
\end{figure}

Figure~\ref{fig:sim_compare} shows the results of the different simulation methods that are described below. We see that both circuit simulations (S-parameter, blue, and harmonic balance, orange) are in close mutual agreement, and both show higher gain and ripple than the ideal coupled-mode simulation (green). This is a consequence of the manual trimming of the electrical length of resonator 3 as was mentioned in the main text, as well as the frequency dependence of all coupling structures.

\subsection{Coupled modes matrix}\label{apx:matrix}
The coupled-mode graph of the LESA circuit is shown in the main text, Fig.~1(a). From the graph, we can write the coupled-mode equations-of-motion matrix $\mathbf{M}$ in the mode basis $\vec{v}=(3, 2, 1, 1^*, 2^*, 3^*)$, as described in Ref.~\cite{naaman2022synthesis},

\begin{equation}\label{eq:PA_matrix}
    \mathbf{M}=
    \begin{bmatrix}
        \Delta_{3} & \beta_{23} & 0 & 0 & 0 & 0\\
        \beta_{23} & \Delta_{2} & \beta_{12} & 0 & 0 & 0\\
        0 & \beta_{12} & \Delta_{1} & \beta_p & 0 & 0\\
        0 & 0 & -\beta^*_p & -\Delta^*_{1} & -\beta_{12} & 0\\
        0 & 0 & 0 & -\beta_{12} & -\Delta^*_{2} & -\beta_{23}\\
        0 & 0 & 0 & 0 & -\beta_{23} & -\Delta^*_{3}
    \end{bmatrix}.
\end{equation}
The diagonal elements of $\mathbf{M}$ are the `detuning' terms, which contain the simulation frequency $\omega$, and are given by
\begin{align*}
    \Delta_3 = -\Delta^*_3& = \frac{1}{\gamma_0}\left(\omega-\omega_0 + i\frac{\gamma_0}{2}\right) \\
    \Delta_{1,2} & = \frac{1}{\gamma_0}\left(\omega-\omega_0\right), 
\end{align*}
where we have assumed that the pump frequency is exactly $\omega_P=2\omega_0$, and $\omega_0$ is the frequency of all the resonant modes in the circuit. The off-diagonal terms are the reduced coupling rates $\beta_{jk}$ for the passive couplers, and $\beta_p$ for the parametric coupler. The values of these terms are given in the main text and calculated based on the prototype coefficients and the bandwidth of the network, $\beta_{23} = 0.339$, $\beta_{12} = 0.27$, $\beta_p = 0.288$. The port dissipation rate is $\gamma_0/2\pi = 1.95$~GHz.

The signal gain of the network $G_s$, measured in reflection off of mode 3, can be calculated \cite{naaman2022synthesis} from the $\left[1,1\right]$ element of the inverse matrix $\mathbf{M}^{-1}$
\begin{equation}\label{eq:matrix_gain}
    \sqrt{G_s} = i\left[\mathbf{M}^{-1}\right]_{1,1} - 1.
\end{equation}
The gain calculated according to Eq.~(\ref{eq:matrix_gain}) is shown in Fig.~\ref{fig:sim_compare} in green. The idler trans-gain $G_i$ can similarly be calculated as the transmission between modes 3 and $3^*$, using the $\left[6,1\right]$ element of the inverse matrix
\begin{equation}
    \sqrt{G_i} = i\left[\mathbf{M}^{-1}\right]_{6,1}.
\end{equation}

\subsection{Harmonic balance}
We have performed circuit simulations of the LESA in Keysight ADS using a harmonic balance simulator with a nonlinear equation-based model for the snake. The implementation details could vary significantly depending on which simulation tool one chooses to use, so we will give here just the basic procedure we have used.

We use the circuit schematic of the LESA in Fig.~1(c) in the main text. The parametrically pumped snake is modeled using a two-port equation-based nonlinear block in ADS. We use one port of the block to represent the signal current and voltage across the snake inductance, and the other port is used to numerically pump the model.

In the simulator, the frequency-domain current $I_s$ and voltage $V_s$ at the `signal' port are evaluated by solving the equation
\begin{equation}\label{eq:hb_model}
    V_s - j\omega I_s\times L_\mathrm{snake}(\delta) = 0
\end{equation}
for each of the harmonics in the problem, where $L_\mathrm{snake}$ is the snake inductance, Eq.~(\ref{snake2}). The phase $\delta$ in turn is represented by a `voltage' $V_p$, measured at the `pump' port of the block. This numerical pump and flux bias are produced in the simulation by a voltage source oscillating at the pump frequency and with a dc component. This model was used in the simulation shown in Fig.~\ref{fig:sim_compare} (orange).

As we can see, this model only approximates the behavior of the snake. First, we are numerically pumping the snake's phase directly instead of pumping a flux bias. Second, the model is linearized, in that the signal current does not affect the snake phase, so that it is inherently in the small-signal limit. 

To go beyond the small-signal approximation, we calculate the phase $\delta_s = (2\pi/\Phi_0)L_\mathrm{snake}(\delta_0)I_s$ associated with the signal current flowing through the unperturbed snake inductance at the dc operating point. We then perturbatively replace $\delta$ in Eq.~(\ref{eq:hb_model}) by $\delta-\delta_s/2N$ in one of the 2 parallel rf-SQUID arrays of the snake and $\delta+\delta_s/2N$ in the other. Doing so we can model the behavior of the amplifier near saturation, including intermodulation distortion.  

\subsection{Linear S-parameter simulation}
Harmonic balance simulations are less straightforward to set up, and are more computationally expensive than S-parameter simulations. Fortunately, if we are only interested in the amplifier's small-signal response, it is possible to simulate it using a linear S-parameter circuit simulation, with a setup that is more standard and more transferable between tools. We describe these simulations here.

\begin{figure*}
\includegraphics[width=\textwidth]{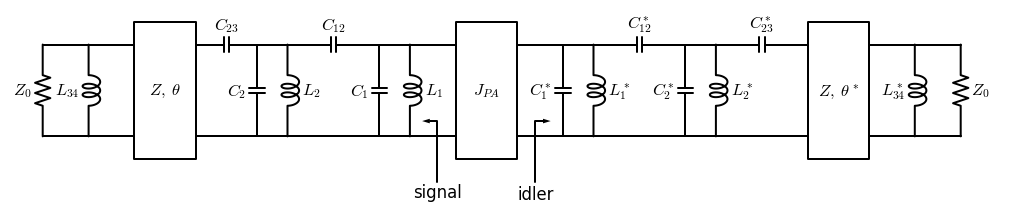}
\caption{\label{fig:sim_schem} Circuit schematic used in linear simulations of the LESA. The idler circuit mirrors the signal circuit, and the starred components are evaluated at the idler frequency $-\omega_i=\omega-\omega_P$. The two circuits are coupled via a parametric admittance inverter block $J_{PA}$.}
\end{figure*}

The circuit schematic used to simulate the LESA is shown in Fig.~\ref{fig:sim_schem}. The circuit is composed of a signal circuit (shown on the left), an idler circuit (shown on the right), and a parametric coupling element, the admittance inverter $J_{PA}$, connecting them. The signal circuit is composed of the linear matching network, including the linear inductance of the snake at the operating point. The inductors, capacitors, and transmission line elements used here are the standard components available in the tool. The idler circuit mirrors the topology and component values of the signal circuit, however, it has to be evaluated at the idler frequency $-\omega_i=\omega_s-\omega_P$, which standard components are not designed to do. The components of the idler circuit, evaluated at $-\omega_i$ are shown as starred in Fig.~\ref{fig:sim_schem}. 

To enable linear S-parameter simulation, we have to define the idler components with their special frequency dependence. In ADS, we use 1-port Equation-Based Linear admittance (impedance) matrix component to define an idler capacitor (inductor). For the idler transmission lines, we use a 2-port $ABCD$ matrix (T-matrix) component.

Taking the simulation frequency variable to be $\omega$, the idler capacitor $C^*$ can be defined as a sub-circuit with parameters $C$ (the capacitance) and $\omega_P$ (the pump frequency) using its admittance matrix
\begin{equation}
    Y[1,1] = j(\omega-\omega_P)C.
\end{equation}
Similarly, the idler inductor $L^*$ can be defined using its impedance matrix
\begin{equation}
    Z[1,1] = j(\omega-\omega_P)L,
\end{equation}
where $L$ is a sub-circuit parameter. The ideal transmission line element (TLIN component in ADS) is defined using its impedance $Z$, electrical length $\theta$, and frequency $\omega_0$. To implement an equivalent idler transmission line, we use the 2-port $ABCD$ matrix 
\begin{equation}
    \mathrm{\mathbf{T}_\mathrm{TLIN}}=
    \begin{bmatrix}
       \cos\left[\left(\omega-\omega_P\right)\tau\right] & jZ\sin\left[\left(\omega-\omega_P\right)\tau\right]\\
       \left(j/Z\right)\sin\left[\left(\omega-\omega_P\right)\tau\right] & \cos\left[\left(\omega-\omega_P\right)\tau\right]
    \end{bmatrix},
\end{equation}
where $\tau=\theta/\omega_0$.

The two circuits are coupled by a parametric admittance inverter \cite{naaman2022synthesis} $J_{PA}$, which we can implement as a 2-port $ABCD$ matrix component
\begin{equation}
    \mathrm{\mathbf{T}_{PA}}=
    \begin{bmatrix}
       0 & j/J_{PA}\\
       -jJ_{PA} & 0
    \end{bmatrix}.
\end{equation}
The value of the admittance inverter $J_{PA}$ can be calculated with
\begin{equation}\label{eq:para_inverter}
   J_{PA} = \frac{w}{Z_1g_1\sqrt{g_0}}\times \sqrt{g_{N+1}^p},
\end{equation}
where $p=+1$ if the order of the matching network $N$ is even and $p=-1$ if $N$ is odd, $w$ is the fractional bandwidth, and $g_k$ are the $k^\mathrm{th}$ prototype coefficients. For example, in our $3^\mathrm{rd}$-order network, we have
\begin{equation}
    J_{PA} = \frac{w}{Z_1g_1\sqrt{g_0g_4}}.
\end{equation}
Alternatively, we can express $J_{PA}$ in terms of the amplifier power gain $G$ for all network orders,
\begin{equation}
    J_{PA}=\frac{w}{Z_1g_1}\left[\frac{\sqrt{G}+\sqrt{G-1}+1}{\sqrt{G}+\sqrt{G-1}-1}\right]^\frac{1}{2}.
\end{equation}

\section{Calculation of circuit components}\label{apx:componenets}

\begin{figure}[b]
\includegraphics[width=\columnwidth]{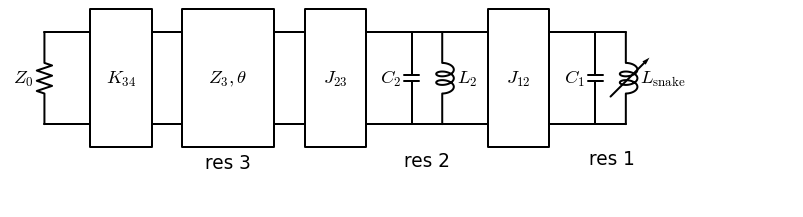}
\caption{\label{fig:block} Block diagram of the LESA matching network. }
\end{figure}

Here, we calculate component values for the LESA matching network. The circuit block diagram is shown in Fig.~\ref{fig:block}, and the schematic is shown in Fig.~1(c) in the main text. As discussed in the main text, the matching network was designed for a center frequency of $\omega_0/2\pi=4.9$~GHz, but the measured amplifiers have a center frequency of $4.6$~GHz. Below we report on the `as designed' parameters, keeping in mind that the `as fabricated' parameters likely differ.

Resonator `res 1' in Fig.~\ref{fig:block}, having a characteristic impedance $Z_1$, is the nonlinear resonator that contains the capacitively-shunted snake element. The inductance $L_\mathrm{snake}$ is given by \cite{white2022readout}
\begin{equation}
L_\mathrm{snake} = L_b+\frac{N}{2} \times \frac{L_J (L^s_1 + L^s_2) + L^s_1 L^s_2 \cos\delta_0}{L_J + (4L^s_1 + L^s_2)\cos\delta_0}, \label{snake2}
\end{equation}
where $L^s_1=2.6$~pH, $L^s_2=8.0$~pH, and $L_J=\hbar/2eI_0$ with $I_0=16\,\mu$A, and $\delta_0$ is the equilibrium junction phase at the operating flux-bias point. $L_b$ is a stray linear inductance associated with the snake wiring, and we assume $L_b=50$~pH.

Resonator `res 2' is a lumped-element parallel $LC$ resonator with characteristic impedance $Z_2$, and `res 3' is a transmission line resonator, having a characteristic impedance $Z_3$ and an electrical length $\theta$ at the center frequency of the amplifier. The resonators are interconnected via admittance inverters, $J_{12}$ and $J_{23}$, and resonator 3 is connected to the $Z_0=50\,\Omega$ signal port via an impedance inverter $K_{34}$.

We start from the network coefficients, using a 20 dB gain, 0.5 dB ripple third-order Chebyshev prototype \cite{naaman2022synthesis} 
\begin{equation}\label{eq:apx_coeffs}
\left\{g_0,\dots,g_4\right\}=\left\{1.0,\; 0.5899,\;0.6681,\;0.3753,\;0.9045\right\},
\end{equation}
and design for a fractional bandwidth of $w=0.135$. We choose an operating point where the snake is biased at about $0.25\Phi_0$ per rf-SQUID stage \cite{white2022readout}, targeting $L_\mathrm{snake}=144$~pH, and therefore $Z_1=\omega_0L_\mathrm{snake}=4.42\,\Omega$. We chose resonator 2 characteristic impedance to be $Z_2=20\,\Omega$, and that of resonator 3 to be $Z_3=50\,\Omega$.

Next, we calculate the values of the immittance inverters \cite{MYJ},
\begin{align}
    J_{12} &= w \sqrt{\frac{1}{g_1g_2Z_1Z_2}} = 0.0228\,\Omega^{-1}, \label{eq:apx_j12}\\
    J_{23} &= w \sqrt{\frac{\pi}{4g_2g_3Z_2Z_3}} = 0.0076\,\Omega^{-1}, \label{eq:apx_j23}\\
    K_{34} &= \sqrt{\frac{\pi}{4}\frac{w Z_3Z_0}{g_3g_4}} = 27.95\,\Omega, \label{eq:apx_k34}
\end{align}
where the $\pi/4$ factors in Eqs.~(\ref{eq:apx_j23}) and~(\ref{eq:apx_k34}) come from resonator 3 being a transmission line quarter wave resonator instead of a lumped element one.

We implement admittance inverter $J_{12}$, disposed between two lumped element resonators, using a series coupling capacitor $C_{12}$, whose value is $C_{12}=J_{12}/\omega_0=0.743$~pF. Impedance inverter $K_{34}$, disposed between two transmission lines (resonator 3 and the 50~$\Omega$ feedline) is implemented as a shunt inductor $L_{34} = X_{34}/\omega_0 = 1.32$~nH, where the reactance $X_{34}$ is given by \cite{MYJ, collin2007foundations}
\begin{equation}
    X_{34}=\frac{K_{34}}{1-\left(K_{34}/Z_3\right)^2}.
\end{equation}

Admittance inverter $J_{23}$ is more unusual, as it is disposed between a lumped-element resonator on one side, and a transmission line resonator on the other side (see Appendix~\ref{apx:inverter}). It is implemented as a series capacitor $C_{23}=B_{23}/\omega_0=0.265$~pF, where 
\begin{equation}
    B_{23}=\frac{J_{23}}{\sqrt{1-\left(J_{23}Z_3\right)^2}}.
\end{equation}

Now that the inverters are calculated, we can calculate the rest of the circuit elements. Capacitor $C_1$ is calculated according to
\begin{equation}
    C_1 = \frac{1}{Z_1\omega_0}-C_{12} = 6.61\,\mathrm{pF}.
\end{equation}
Resonator 2 components are
\begin{align}
    L_2 &= \frac{Z_2}{\omega_0} = 0.65\,\mathrm{nH} \\
    C_2 &= \frac{1}{Z_2\omega_0} - C_{12}-B_{23e}/\omega_0 = 0.654\,\mathrm{pF},
\end{align}
where $B_{23e}=J_{23}\sqrt{1-\left(J_{23}Z_3\right)^2}$. Finally, resonator 3 electrical length is given by
\begin{align}
    \theta &= \frac{\pi}{2} - \tan^{-1}\left(B_{23}Z_3\right)-\frac{1}{2}\tan^{-1}\left(2X_{34}/Z_3\right)\nonumber\\
    &=38.6^\circ.
\end{align}
This length was further trimmed manually to $\theta=32.6^\circ$ as described in the main text.

\section{Admittance inverter between a lumped and a transmission line resonator}\label{apx:inverter}

The usual literature has examples and design equations for admittance inverters disposed between same-type resonators \cite{pozar2009microwave} (lumped or transmission line). In the present circuit, we would like to implement an inverter that has a lumped element resonator on one side, and a quarter-wave transmission line resonator on the other side. We do not know of an easily accessible example of this case in the literature, so we will derive the design equations here.

\begin{figure}
\includegraphics[width=\columnwidth]{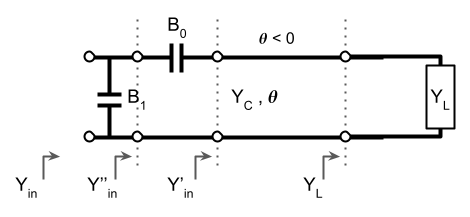}
\caption{\label{fig:inverter} Schematic of the inverter circuit, containing a series coupling capacitor $B_0$, a compensating (negative) shunt capacitor $B_1$, and a compensating (negative) transmission line length with admittance $Y_c$ and electrical length $\theta$. }
\end{figure}

We will follow a procedure similar to that described by Collin \cite{collin2007foundations}. The plan is to calculate $Y_\mathrm{in}$, the input admittance seen from the left side of Fig.~\ref{fig:inverter}. If the circuit is to function as an admittance inverter $J$, we should have $Y_\mathrm{in}=\frac{J^2}{Y_\mathrm{L}}$. Given $J$ and the admittance of the transmission line $Y_\mathrm{c}$, we will calculate the susceptances $B_0$, $B_1$ and the (negative) electrical length $\theta$. Susceptance $B_0$ will relate to the coupling capacitor (via $B_{23}$ in Section~\ref{apx:componenets}), susceptance $B_1$ will be absorbed into the lumped element resonator on one side of the inverter ($B_{23e}$ in Section~\ref{apx:componenets}), and $\theta<0$ will be used to compensate the transmission line resonator on the other side of the inverter.

We start by calculating $Y'_\mathrm{in}$, looking from the left of the transmission line into the load in Fig.~\ref{fig:inverter}:
\begin{equation}\label{eq:y_prime}
    Y'_\mathrm{in}=Y_\mathrm{c}\frac{Y_\mathrm{L}-jY_\mathrm{c}t}{Y_\mathrm{c}-jY_\mathrm{L}t},
\end{equation}
where $t=\tan|\theta|$. Next, the admittance $Y''_\mathrm{in}$ looking from the left of the susceptance $B_0$ is given by:
\begin{equation}\label{eq:y_double_prime}
    Y''_\mathrm{in}=\frac{jB_0Y'_\mathrm{in}}{jB_0+Y'_\mathrm{in}}.
\end{equation}
Finally, the input admittance is
\begin{equation}\label{eq:y_in}
    Y_\mathrm{in} = jB_1+Y''_\mathrm{in}=jB_1+\frac{jB_0Y'_\mathrm{in}}{jB_0+Y'_\mathrm{in}},
\end{equation}
where we have used Eq.~(\ref{eq:y_double_prime}). Further plugging in $Y'_\mathrm{in}$ from Eq.~(\ref{eq:y_prime}) and some algebra, we get:
\begin{align}\label{eq:y_in_long}
    Y_\mathrm{in}&= \nonumber\\
    &\frac{jY_\mathrm{L}\left[ B_0B_1+\left(B_0+B_1\right)Y_\mathrm{c}\right]-B_0B_1Y_\mathrm{c}+\left(B_0+B_1\right)Y_\mathrm{c}^2t}{Y_\mathrm{L}\left[B_0t+Y_\mathrm{c}\right]+jY_\mathrm{c}\left[B_0-Y_\mathrm{c}t\right]}
\end{align}

Next we want to bring this to the form $J^2/Y_\mathrm{L}$, so we see that we need to zero out the $jY_\mathrm{L}$ term in the numerator, and zero out the $jY_\mathrm{c}$ term in the denominator. These two conditions are satisfied with:
\begin{equation}\label{eq:b0}
    B_0 = Y_\mathrm{c}t
\end{equation}
and
\begin{equation}\label{eq:b1}
    B_1 = -\frac{B_0}{t^2+1}
\end{equation}
Plugging these into Eq.~(\ref{eq:y_in_long}), we finally get after algebra:
\begin{equation}
    Y_\mathrm{in}=\frac{1}{Y_\mathrm{L}}\left[\frac{Y_\mathrm{c}\tan|\theta|}{\sqrt{1+\tan^2|\theta|}}\right]^2,
\end{equation}
from which we can identify the inverter value $J$:
\begin{equation}\label{eq:J}
    J = \frac{Y_\mathrm{c}\tan|\theta|}{\sqrt{1+\tan^2|\theta|}}.
\end{equation}
Using Eq.~(\ref{eq:b0}) and Eq.~(\ref{eq:J}) we can express $B_0$ in terms of $J$:
\begin{equation}
    B_0=\frac{J}{\sqrt{1-\left(J/Y_\mathrm{c}\right)^2}},
\end{equation}
and with Eq.~(\ref{eq:b1}) we express $B_1$ in terms of $J$:
\begin{equation}
    B_1 = -J\times\sqrt{1-\left(J/Y_\mathrm{c}\right)^2},
\end{equation}
and finally, the length of the transmission line $\theta$ in terms of $B_0$:
\begin{equation}
    \theta = -\tan^{-1}\left(\frac{B_0}{Y_\mathrm{c}}\right).
\end{equation}
with the values of $B_0$ and $B_1$ we can calculate the capacitances in Fig.~\ref{fig:inverter}:
\begin{align}
    C_0 & = B_0/\omega_0 \\
    C_1 & = B_1/\omega_0.
\end{align}

\section{Theory of intermodulation distortion due to a saturable TLS bath}\label{apx:imd_thy}

We consider a mode of a nonlinear resonator with coordinate $q$, momentum $p$, and eigenfrequency $\omega_0$. The mode is parametrically pumped at frequency $\omega_p\approx 2\omega_0$, and is driven by a force $F_d(t)$, which we associate with the signal. In addition, the mode is coupled to a bath of two-level systems (TLSs). An $n^\mathrm{th}$ TLS is described by the Pauli operators $\sigma_i\sn$  ($i=x,y,z$) and has transition frequency $\omega\sn$. The Hamiltonian of the system reads \cite{bachtold2022mesoscopic}
\begin{align}
\label{eq:hamiltonian}
    &H = H_\mathrm{res}+\frac{1}{2}q^2F_p\cos(\omega_pt)-qF_d(t) + H_\mathrm{TLS}+H_i\\
    &H_\mathrm{res} =\frac{1}{2}\left(p^2+\omega_0^2q^2\right)+\frac{1}{4}\gamma q^4\\
    &H_\mathrm{TLS} = -\frac{1}{2}\sum_n{\hbar\omega^{(n)}\sigma_z}\\
    &H_i = -\sum_n{v^{(n)}q\sigma_x}
\end{align}
Here $\gamma$ is the parameter of the mode nonlinearity (the high-frequency Kerr coefficient) and $v^{(n)}$ is the parameter of the coupling of the mode to the $n^\mathrm{th}$ TLS. For the considered electromagnetic mode, this parameter is determined by the dipole moment of the TLS.
We consider coupling to resonant TLSs, $\omega\sn$ close to $\omega_0$.

The coupling of the TLSs to excitations in the material, in particular to phonons, leads to decay of the TLSs. In turn, this creates a decay channel, which we call ``internal'' decay. The coupling of the resonator, through the LESA matching circuit to the $50\,\Omega$ environment, also leads to mode decay, which we call ``external''. If the total decay rate of the mode is $\kappa$, the linear susceptibility of the system with respect to a signal at frequency $\omega$ is
\begin{align}
\label{eq:susceptibility_kappa}
    &\chi(\omega)=\frac{i}{\omega_p}\frac{\kappa-i(\omega +\omega_0 -\omega_p)}{[\kappa-i(\omega-\omega_p/2)]^2-\kappa^2(f_p^2-\mu_p^2)}, \nonumber\\
    &f_p = F_p/2\kappa\omega_p, \quad \mu_p = (\omega_p-2\omega_0)/2\kappa.
\end{align}
Equation (\ref{eq:susceptibility_kappa}) is written for the case where both the signal frequency and half the parametric pump frequency are close to the mode eigenfrequency, $|\omega_p/2-\omega_0|,\, |\omega-\omega_0|\ll \omega_0$. We also assumed that the mode decay rate is comparatively small, $\kappa\ll \omega_0$. The parameter $f_p= F_p/2\kappa\omega_p$ is the scaled strength of the pump.  

Of interest for the experiment is resonant pumping, $\omega_p=2\omega_0$, in which case $\mu_p=0$.  Here, for a signal sharp on resonance, $\omega=\omega_0$,
\begin{equation}
\label{eq:suscept}
\chi(\omega_0)=i\sqrt{G}/2\omega_0\kappa  \qquad (\omega_p=2\omega_0),
\end{equation}
where $G=(1-f_p^2)^{-2}$ is the amplifier power gain.

We can relate quantities appearing in Eq.~(\ref{eq:hamiltonian}) to experimentally accessible ones by thinking of the resonator as an $LC$ resonator coupled to a transmission line; such resonator models the primary resonator of the LESA. The coordinate $q$ relates to the voltage $V$ via $q=V/\omega_0^{3/2}Z_r^{1/2}$, where $Z_r$ is the impedance of the resonator. The drive force $F_d$ relates to the drive voltage $V_d$ via $F_d=2\kappa' V_d/\sqrt{\omega_0Z_r}$; here $\kappa'$ characterizes the external decay due to coupling to the environment; in the experimentally studied system it is close to the total decay rate $\kappa$. Since the resonator  is embedded in a matching network, we can relate $V_d$ to the voltage $V_{in}$ at the input terminal of the LESA,  $V_d=V_{in}\sqrt{g_1Z_r/wZ_0}$, where $g_1$ is the network prototype coefficient, $Z_0=50\,\Omega$ is the environment impedance, and $w$ is the network fractional bandwidth.

When the resonator is driven by two tones at frequencies $\omega_1$ and $\omega_2$ of equal amplitude $F_c$, which are centered at $\omega_c= (\omega_1 + \omega_2)/2$ and spaced by $\Delta\omega = \omega_1 - \omega_2$, the driving $F_d(t)$ can be written as
\begin{align}
    \label{eq:combined_drive}
    F_d(t)=F_ce^{-i\omega_ct}\cos{\left(\Delta\omega\,t/2\right)}+ \mathrm{c.c.}.
\end{align}
We study the nonlinear response to this driving for $\omega_c$ close to the mode eigenfrequency $\omega_0$ and $|\Delta\omega| \ll \omega_0$. It is convenient to analyze this response by switching to the complex amplitude of the mode $a(t)$ that varies slowly on the time scale $\omega_p^{-1}$,
\begin{align}
\label{eq:complex_amplitude}
a(t) = \frac{1}{2}\left(q+ i\frac{2p}{\omega_p}\right) e^{i\omega_pt/2}. 
\end{align}
If we disregard the mode nonlinearity and the coupling to the TLSs, we have in the rotating wave approximation
\begin{align}
\label{eq:lin_response}
a_\mathrm{lin}(t)= \chi(\omega_c)F_c e^{-i(\omega_c-\omega_p/2)t}\cos{\left(\Delta\omega\,t/2\right)}.
\end{align}

The further analysis is based on the following picture. The TLSs are coupled to the driving via their coupling to the mode. In turn, their response affects the mode itself. This response becomes nonlinear well before the Kerr nonlinearity comes into play. As a result, the response of the mode to the drive also becomes nonlinear. However, we will assume that the overall nonlinearity of the mode dynamics (but not the TLS dynamics) is weak. Therefore in the analysis of the TLS dynamics one can approximate the mode dynamics by Eqs.~(\ref{eq:complex_amplitude}) and (\ref{eq:lin_response}).

To study the TLS dynamics we go to the rotating frame  using the standard transformation $U(t) = \prod_n\exp(i\omega_p t\sigma_z\sn/4)$. Then the TLS operators $\sigma_\pm\sn =\sigma_x\sn \pm i \sigma_y\sn$ take the form $ \sigma_\pm\sn (t) = \exp(\mp i \omega_pt/2)\tilde\sigma_\pm\sn(t)$, where $\tilde\sigma_\pm\sn(t)$ are slowly varying on the time scale $\omega_p^{-1}$. As we will see, the drive (\ref{eq:combined_drive}) makes $\tilde\sigma\sn$ oscillate at frequencies $|(k+1)\omega_1 - k\omega_2| - \omega_p/2$ (with integer $k$).  
In the rotating wave approximation the effect of these oscillations of the TLSs on $a(t)$ is described by the expression 
\begin{equation}
\label{eq:q_tls_sum}
    a_\mathrm{TLS}(t) =\frac{1}{2}\chi(\omega_c)\sum_n v^{(n)}\tilde\sigma_+\sn (t).
\end{equation}

We can solve the Bloch equations for the TLSs assuming that, in the coupling Hamiltonian $H_i$, $q(t) = a_\mathrm{lin}(t)\exp(-i\omega_pt/2) + \mathrm{c.c.}$. This gives 
\begin{align}
\label{eq:polarization}
    \tilde\sigma_+^{(n)}&=\frac{2iv^{(n)}T_2\sn}{\hbar}\chi(\omega_c) F_c  e^{-i(\omega_c - \omega_p/2)t}\cos(\Delta\omega t/2)\nonumber\\
    &\times\frac{1+iT_2\sn\delta\omega\sn}{1+\left(T_2\sn\delta\omega\sn\right)^2 +\zeta\sn\left[1+\cos\left(\Delta\omega\,t\right)\right]}.
\end{align}
Here $\delta\omega\sn = \omega_c - \omega\sn$ is the detuning of the resonant frequency of the $n^\mathrm{th}$ TLS away from $\omega_c$, whereas $T_1\sn$ and $T_2\sn$ are its decay and decoherence times.  The dimensionless parameter $\zeta^{(n)}$ is 
\begin{equation}
\label{eq:zeta_sn}
    \zeta^{(n)} = \hbar^{-2}T_1\sn T_2\sn |v^{(n)}|^2 A^2,\quad A=\sqrt{2}|\chi(\omega_c)|F_c.
\end{equation}
The parameter $A$ is the amplitude of the 2-tone beat envelope in the linear approximation given by Eq.~(\ref{eq:lin_response}). 

Equation (\ref{eq:polarization}) applies provided the difference between the tone frequencies $|\Delta\omega|$ is small compared to the relaxation rates of the relevant TLSs $1/T_1\sn, 1/T_2\sn$. This allowed us to assume that the TLSs follow the oscillations of $a_\mathrm{lin}(t)$ adiabatically, i.e., to disregard delay in describing the response of the TLSs to the two-tone drive.

By Fourier-expanding $\tilde\sigma_+\sn$ in a  series in $\exp(i\Delta\omega t)$ one finds from Eqs.~(\ref{eq:q_tls_sum}) and (\ref{eq:polarization}) that $\tilde\sigma_+\sn$, and thus $a_\mathrm{TLS}$, are sums of terms $\tilde\sigma\sn[k], a_\mathrm{TLS}[k]$ that oscillate at the combination frequencies, i.e., 
\begin{align}
\label{eq:kth_components}
&a_\mathrm{TLS}[k]\propto \tilde\sigma\sn[k] \propto \exp(-i\delta\Omega[k]t), \nonumber\\
&\delta\Omega[k]=(k+1)\omega_1- k\omega_2 -\omega_p/2
\end{align}
with integer $k\neq 0,-1$. From Eq.~(\ref{eq:complex_amplitude}), this corresponds to the  mode vibrations at frequencies $(k+1)\omega_1- k\omega_2$. Such vibrations describe the intermodulation due to the coupling to the TLSs. 

The amplitude of the vibrations at frequencies $(k+1)\omega_1- k\omega_2$ is determined by the   parameter $\zeta\sn$. This parameter can be large even where the Kerr nonlinearity is still small. We note that $\Omega_R\sn = |v\sn| A/\hbar$  can be thought of as the Rabi frequency of the $n^\mathrm{th}$ TLS in response to the ``drive'' with amplitude $A$ at frequency $\omega_c$. Therefore $\zeta\sn$ has a familiar form of $T_1\sn T_2\sn\Omega_R\sn{}^2$. 

Further simplification of the general expressions for the intermodulation amplitudes can be made by assuming that the TLSs are dipoles with random orientation and there is no correlation between this orientation and other parameters of the TLS, i.e., one can set 
\[v\sn = V\sn\cos\theta\sn,\]
where $\theta\sn$ is the random angle between the dipole moment and the mode field. 

\subsection{Third order product}
On averaging over $\theta\sn$ one obtains, for the vibrations at frequency $2\omega_1-\omega_2$,
\begin{align}
\label{eq:q_IM_general}
&q_\mathrm{TLS}^{2\omega_1-\omega_2} =   a_\mathrm{TLS}^{2\omega_1-\omega_2}
e^{-i(2\omega_1-\omega_2)t} + \mathrm{c.c.}, \nonumber\\
& a_\mathrm{TLS}^{2\omega_1-\omega_2} = \braket{a_\mathrm{TLS}[1]}_{\{\theta\sn\}} e^{i(2\omega_1-\omega_2-\omega_p/2)t}.
\end{align}
Here $a_\mathrm{TLS}[1]$ is the component of $a_\mathrm{TLS}$ that is $\propto \exp[-i\delta\Omega[1] t]$, cf. Eq.~(\ref{eq:kth_components}), and $\braket{...}_{\{\theta\sn\}}$ indicates averaging over the angles $\theta\sn$.

From Eqs. (\ref{eq:q_tls_sum}) and (\ref{eq:polarization})  we find
\begin{align}
\label{eq:ampl_before_omega_aver}
&a_\mathrm{TLS}^{2\omega_1-\omega_2}=-i\frac{\chi^2(\omega_c)}{F_c\,|\chi(\omega_c)|^2}\sum_n\frac{\hbar}{4T_1\sn}\,\Psi^{2\omega_1-\omega_2}(\xi\sn),\nonumber\\
&\Psi^{2\omega_1-\omega_2}(\xi) = 
 \frac{3}{4}\xi^{-1/2}\,\log\left(\sqrt{\xi} + \sqrt{\xi + 1}\right)\nonumber\\
&+\frac{1}{4}\,(\xi +1)^{1/2}  -1
\end{align}
where
\begin{align}
\label{eq:xi_zeta_sn}
&\xi\sn = 2\bar\zeta\sn/[1+(T_2\sn\delta\omega\sn)^2], \nonumber\\ 
&\bar\zeta\sn = \hbar^{-2}T_1\sn T_2\sn |V^{(n)}|^2 A^2,
\end{align}

We assume that there are many resonant TLSs. Then the sum over $n$ in Eq.~(\ref{eq:ampl_before_omega_aver}) involves averaging over the TLSs. We will do this averaging in the common assumption that the relaxation rates of different TLSs are approximately the same as are also their effective dipole moments $V\sn$, while the major randomness comes from the distribution of the TLS eigenfrequencies $\omega\sn$ \cite{phillips1987two}. In this approximation the values of $\bar\zeta\sn$ are the same, 
\begin{equation}
\label{eq:zeta}
\bar\zeta\sn=   \bar \zeta=T_1T_2\bar{\Omega}_R^2,\quad \bar\Omega_R = VA/\hbar, 
\end{equation}
identifying $\bar{\Omega}_R$ as the effective Rabi frequency under the 2-tone drive. 

We carry out the sum in Eq.~(\ref{eq:ampl_before_omega_aver}) by integrating over the TLS detuning $\delta\omega\sn$ with the weighting factor $\rho$, which is determined by the number of TLSs per unit bandwidth, 
\begin{align}
\label{eq:mean_Psi}
&\sum_n\Psi^{2\omega_1-\omega_2}(\xi\sn)/T_1\sn=\rho \langle \Psi^{2\omega_1-\omega_2} \rangle/T_1, \nonumber\\
&\langle \Psi^{2\omega_1-\omega_2} \rangle=\int d\omega\sn \Psi^{2\omega_1-\omega_2}(\bar\xi\sn), 
\end{align}
where $\bar\xi\sn = 2\bar\zeta/[1+ (T_2\delta\omega\sn)^2]$. In view of this averaging we dropped the term $\propto \delta\omega\sn$ in $a_\mathrm{TLS}^{2\omega_1-\omega_2}$ that comes from the odd term $iT_2\sn\delta\omega\sn$ in the numerator in Eq.~(\ref{eq:polarization}). 

The integral, Eq.~(\ref{eq:mean_Psi}), can be evaluated numerically. Since the amplifier is wide band, we approximate the susceptibility using Eq.~(\ref{eq:suscept}) and simplify,
\begin{equation}\label{eq:q_tls_simply}
    a_\mathrm{TLS}^{2\omega_1-\omega_2} =iF_c\frac{G}{8\omega_0^2\kappa^2}\frac{\rho V^2}{\hbar T_1\bar{\Omega}_R^2}\langle\Psi^{2\omega_1-\omega_2}\rangle.
\end{equation}

The quantity $\rho V^2$ can be estimated from the mode decay rate $\Gamma_\mathrm{TLS} = \kappa-\kappa'$ due to the unsaturated TLS bath at low drive powers
\begin{equation}
\label{eq:rhoV2}
    \rho V^2=\frac{6}{\pi}\hbar\omega_0\Gamma_\mathrm{TLS}=\frac{3}{\pi}\frac{\hbar\omega_0^2}{Q_i},
\end{equation}
where $Q_i$ is the internal quality factor of the LESA resonator. Combining Eqs.~(\ref{eq:q_tls_simply}) and (\ref{eq:rhoV2}), we finally get
\begin{equation}
\label{eq:a_2omega_via_G}
    a_\mathrm{TLS}^{2\omega_1-\omega_2}=i\frac{3G}{8\pi Q_i}\frac{F_c}{T_1\kappa^2\bar{\Omega}_R^2}\langle\Psi^{2\omega_1-\omega_2}\rangle,
\end{equation}
and converting to voltages we arrive at 
\begin{equation}
    V_\mathrm{TLS} = i\frac{3G}{4\pi Q_i}\frac{\omega_0V_d}{\kappa T_1\bar{\Omega}_R^2(V_d)}\langle \Psi^{2\omega_1-\omega_2} \rangle,
\end{equation}
which is Eq.~(\ref{eq:vtls}) up to an overall phase that we suppressed in the main text. The Kerr nonlinearity term in Eq.~(\ref{eq:thy_IMD}) has the same overall phase, but opposite sign.

\begin{figure}
\includegraphics[width=\columnwidth]{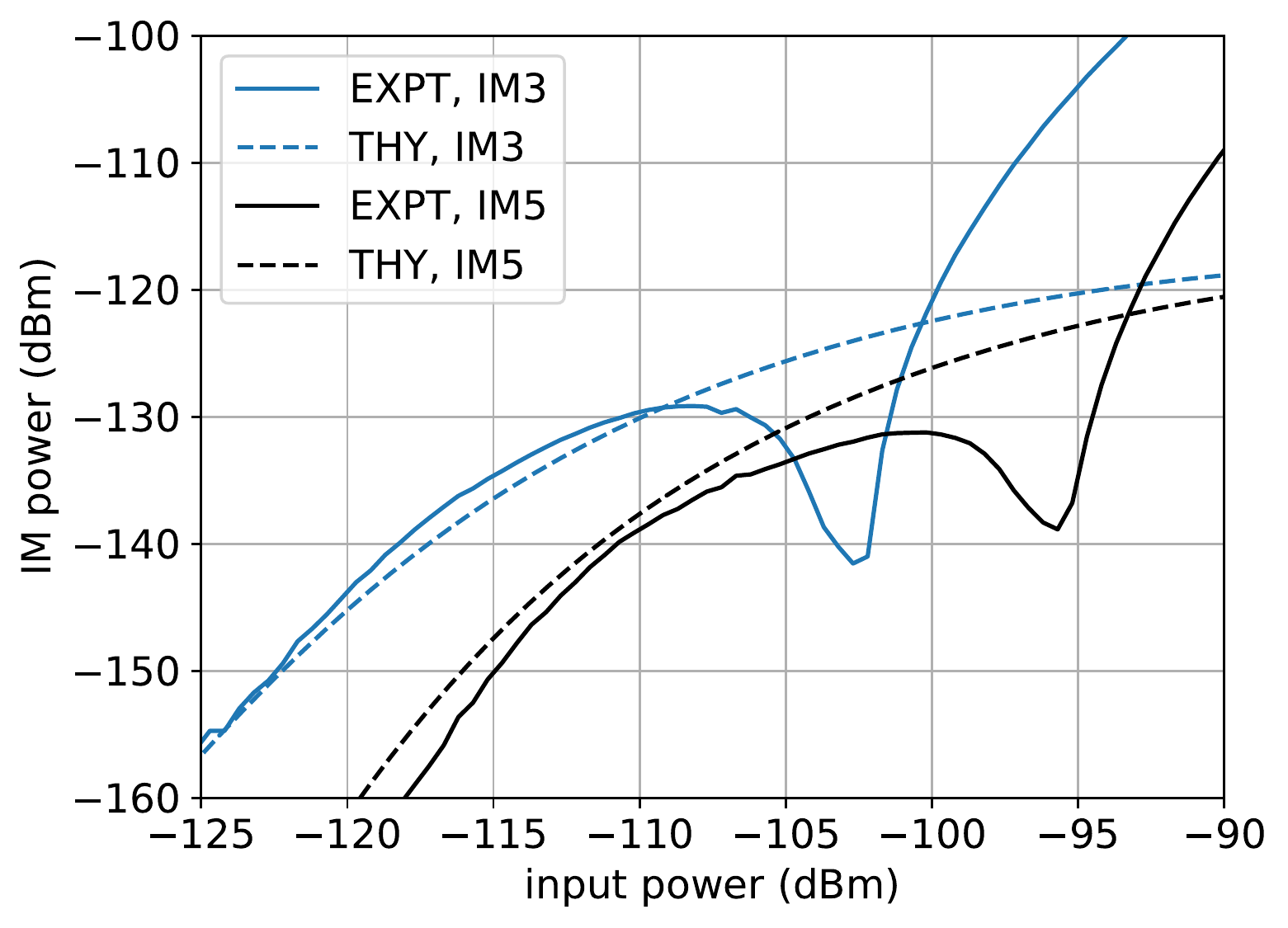}
\caption{\label{fig:IMD35} 3rd order (blue) and 5th (black) order IM product, measured at $\Delta\omega/2\pi=10$~kHz inter-tone detuning (solid) and compared to the TLS contribution from theory (dashed), using the same parameters as in Fig.~\ref{fig:TLS}.}
\end{figure}

\subsection{Fifth order product}
The calculation for the fifth-order intermodulation product at $3\omega_1-2\omega_2$ is exactly the same as for mixing at $2\omega_1 - \omega_2$. We have
\begin{align}
\label{eq:q_3omega_1}
&q_\mathrm{TLS}^{3\omega_1-2\omega_2} =   a_\mathrm{TLS}^{3\omega_1-2\omega_2}
e^{-i(3\omega_1-2\omega_2)t} + \mathrm{c.c.}, \nonumber\\
& a_\mathrm{TLS}^{3\omega_1-2\omega_2} = \braket{a_\mathrm{TLS}[2]}_{\{\theta\sn\}} e^{i(3\omega_1-2\omega_2-\omega_p/2)t}.
\end{align}
Here $a_\mathrm{TLS}[2]$ is the component of $a_\mathrm{TLS}$ that is $\propto \exp[-i\delta\Omega(2) t]$, cf. Eq.~(\ref{eq:kth_components}), and as before, $\braket{...}_{\{\theta\sn\}}$ indicates averaging over the angles $\theta\sn$.

From Eqs. (\ref{eq:q_tls_sum}) and (\ref{eq:polarization})  we find, upon averaging over the dipole orientation,
\begin{align}
\label{eq:ampl_3omega_before_omega_aver}
&a_\mathrm{TLS}^{3\omega_1-2\omega_2}=-i\frac{\chi^2(\omega_c)}{F_c\,|\chi(\omega_c)|^2}\sum_n\frac{\hbar}{4T_1\sn}\,\Psi^{3\omega_1-2\omega_2}(\xi\sn),\nonumber\\
&\Psi^{3\omega_1-2\omega_2}(\xi) = 
 \{16 - 8 \xi +(\xi - 16) \sqrt{1 + \xi}\nonumber\\
 &  + 
 15 \sqrt{\xi}\log[\sqrt{\xi} + \sqrt{\xi +1}]\}/4\xi
\end{align}
where $\xi\sn$ and $\bar \zeta\sn$ are given by Eq.~(\ref{eq:xi_zeta_sn}). For completeness, if we disregard the difference between the values of the dipole moments and the relaxation times of different TLSs, the subsequent averaging over the TLSs frequencies is done exactly in the same way as for the tone at frequency $2\omega_1 - \omega_2$, giving
\begin{equation}
\label{eq:a_3omega_via_G}
    a_\mathrm{TLS}^{3\omega_1-2\omega_2}=i\frac{3G}{8\pi Q_i}\frac{F_c}{T_1\kappa^2\bar{\Omega}_R^2}\langle\Psi^{3\omega_1-2\omega_2}\rangle.
\end{equation}

Figure~\ref{fig:IMD35} shows a comparison of the experimentally measured 3rd and 5th order IM product at $\Delta\omega/2\pi=10$~kHz detuning (solid), compared to the calculated TLS contribution (dashed) based on Eq.~(\ref{eq:a_2omega_via_G}) and~(\ref{eq:a_3omega_via_G}) respectively. The 3rd order intermodulation product is shown in blue, and the 5th order product is in black. We use the same parameters as in Fig.~\ref{fig:TLS} in both equations, and the apparent agreement gives us confidence that the theory captures the essential aspects of this effect.

The above analysis directly extends to a system of several modes with close eigenfrequencies, which are coupled directly and via parametric drive and which are driven by two drives with equal amplitudes and close frequencies. The analysis of intermodulation should take into account that the same TLSs can be coupled to several modes. The results depend on the mode frequencies and coupling, but qualitatively they are similar to those for a single-mode system.

\end{appendix}

%

\end{document}